\documentclass[12pt,twoside,fleqn]{report}

\usepackage{epsfig}
\usepackage{graphics}
\usepackage{graphicx}
\usepackage{subfigure}
\usepackage{exscale}
\usepackage{latexsym}
\usepackage{makeidx}
\usepackage[section]{placeins}
\usepackage{xtab}
\usepackage{cite}

\begin{document}

\newcounter{univ_counter}
\setcounter{univ_counter} {0}
\addtocounter{univ_counter} {1} 
\edef\INFNGE{$^{\arabic{univ_counter}}$ } \addtocounter{univ_counter} {1} 
\edef\MOSCOW{$^{\arabic{univ_counter}}$ } \addtocounter{univ_counter} {1} 
\edef\ROMATRE{$^{\arabic{univ_counter}}$ } \addtocounter{univ_counter} {1} 
\edef\ASU{$^{\arabic{univ_counter}}$ } \addtocounter{univ_counter} {1} 
\edef\UCLA{$^{\arabic{univ_counter}}$ } \addtocounter{univ_counter} {1} 
\edef\CMU{$^{\arabic{univ_counter}}$ } \addtocounter{univ_counter} {1} 
\edef\CUA{$^{\arabic{univ_counter}}$ } \addtocounter{univ_counter} {1} 
\edef\SACLAY{$^{\arabic{univ_counter}}$ } \addtocounter{univ_counter} {1} 
\edef\CNU{$^{\arabic{univ_counter}}$ } \addtocounter{univ_counter} {1} 
\edef\UCONN{$^{\arabic{univ_counter}}$ } \addtocounter{univ_counter} {1} 
\edef\ECOSSEE{$^{\arabic{univ_counter}}$ } \addtocounter{univ_counter} {1} 
\edef\EMMY{$^{\arabic{univ_counter}}$ } \addtocounter{univ_counter} {1} 
\edef\FIU{$^{\arabic{univ_counter}}$ } \addtocounter{univ_counter} {1} 
\edef\FSU{$^{\arabic{univ_counter}}$ } \addtocounter{univ_counter} {1} 
\edef\GWU{$^{\arabic{univ_counter}}$ } \addtocounter{univ_counter} {1} 
\edef\ECOSSEG{$^{\arabic{univ_counter}}$ } \addtocounter{univ_counter} {1} 
\edef\ISU{$^{\arabic{univ_counter}}$ } \addtocounter{univ_counter} {1} 
\edef\INFNFR{$^{\arabic{univ_counter}}$ } \addtocounter{univ_counter} {1} 
\edef\ORSAY{$^{\arabic{univ_counter}}$ } \addtocounter{univ_counter} {1} 
\edef\BONN{$^{\arabic{univ_counter}}$ } \addtocounter{univ_counter} {1} 
\edef\ITEP{$^{\arabic{univ_counter}}$ } \addtocounter{univ_counter} {1} 
\edef\JMU{$^{\arabic{univ_counter}}$ } \addtocounter{univ_counter} {1} 
\edef\KYUNGPOOK{$^{\arabic{univ_counter}}$ } \addtocounter{univ_counter} {1} 
\edef\MIT{$^{\arabic{univ_counter}}$ } \addtocounter{univ_counter} {1} 
\edef\UMASS{$^{\arabic{univ_counter}}$ } \addtocounter{univ_counter} {1} 
\edef\UNH{$^{\arabic{univ_counter}}$ } \addtocounter{univ_counter} {1} 
\edef\NSU{$^{\arabic{univ_counter}}$ } \addtocounter{univ_counter} {1} 
\edef\OHIOU{$^{\arabic{univ_counter}}$ } \addtocounter{univ_counter} {1} 
\edef\ODU{$^{\arabic{univ_counter}}$ } \addtocounter{univ_counter} {1} 
\edef\PITT{$^{\arabic{univ_counter}}$ } \addtocounter{univ_counter} {1} 
\edef\RPI{$^{\arabic{univ_counter}}$ } \addtocounter{univ_counter} {1} 
\edef\RICE{$^{\arabic{univ_counter}}$ } \addtocounter{univ_counter} {1} 
\edef\URICH{$^{\arabic{univ_counter}}$ } \addtocounter{univ_counter} {1} 
\edef\SCAROLINA{$^{\arabic{univ_counter}}$ } \addtocounter{univ_counter} {1} 
\edef\JLAB{$^{\arabic{univ_counter}}$ } \addtocounter{univ_counter} {1} 
\edef\UNIONC{$^{\arabic{univ_counter}}$ } \addtocounter{univ_counter} {1} 
\edef\VT{$^{\arabic{univ_counter}}$ } \addtocounter{univ_counter} {1} 
\edef\VIRGINIA{$^{\arabic{univ_counter}}$ } \addtocounter{univ_counter} {1} 
\edef\WM{$^{\arabic{univ_counter}}$ } \addtocounter{univ_counter} {1} 
\edef\YEREVAN{$^{\arabic{univ_counter}}$ } \addtocounter{univ_counter} {1} 
\edef\deceased{$^{\arabic{univ_counter}}$ } \addtocounter{univ_counter} {1} 
\edef\NOWOHIOU{$^{\arabic{univ_counter}}$ } \addtocounter{univ_counter} {1} 
\edef\NOWINDSTRA{$^{\arabic{univ_counter}}$ } \addtocounter{univ_counter} {1} 
\edef\NOWUNH{$^{\arabic{univ_counter}}$ } \addtocounter{univ_counter} {1} 
\edef\NOWMOSCOW{$^{\arabic{univ_counter}}$ } \addtocounter{univ_counter} {1} 
\edef\NOWSCAROLINA{$^{\arabic{univ_counter}}$ } \addtocounter{univ_counter} {1} 
\edef\NOWUMASS{$^{\arabic{univ_counter}}$ } \addtocounter{univ_counter} {1} 
\edef\NOWMIT{$^{\arabic{univ_counter}}$ } \addtocounter{univ_counter} {1} 
\edef\NOWURICH{$^{\arabic{univ_counter}}$ } \addtocounter{univ_counter} {1} 
\edef\NOWODU{$^{\arabic{univ_counter}}$ } \addtocounter{univ_counter} {1} 
\edef\NOWGEISSEN{$^{\arabic{univ_counter}}$ } \addtocounter{univ_counter} {1} 
\edef\NOWNONE{$^{\arabic{univ_counter}}$ } \addtocounter{univ_counter} {1} 

\begin{titlepage}

\begin{flushright}
CLAS-NOTE 2005-013
\end{flushright}

\vspace{0.3 in}

\begin{center}
{\LARGE\bf The deuteron structure function $F_2$ with CLAS}

\vspace{0.3 in}
\renewcommand{\thefootnote}{\fnsymbol{footnote}}

{\small
M.~Osipenko,\INFNGE$^,$\MOSCOW\
G.~Ricco,\INFNGE\
S.~Simula,\ROMATRE\
M.~Battaglieri,\INFNGE\
M.~Ripani,\INFNGE\
G.~Adams,\RPI\
P.~Ambrozewicz,\FIU\
M.~Anghinolfi,\INFNGE\
B.~Asavapibhop,\UMASS\
G.~Asryan,\YEREVAN
% ...
\
\\{\Large\bf (The CLAS Collaboration)}}

%
\vspace{0.5 in}

{\INFNGE \small INFN, Sezione di Genova, 16146 Genova, Italy} \\
{\MOSCOW \small Moscow State University, General Nuclear Physics Institute, 119899 Moscow, Russia} \\
{\ROMATRE \small Universit\`{a} di ROMA III, 00146 Roma, Italy} \\
{\ASU \small Arizona State University, Tempe, Arizona 85287-1504} \\
{\UCLA \small University of California at Los Angeles, Los Angeles, California  90095-1547} \\
{\CMU \small Carnegie Mellon University, Pittsburgh, Pennsylvania 15213} \\
{\CUA \small Catholic University of America, Washington, D.C. 20064} \\
{\SACLAY \small CEA-Saclay, Service de Physique Nucl\'eaire, F91191 Gif-sur-Yvette,Cedex, France} \\
{\CNU \small Christopher Newport University, Newport News, Virginia 23606} \\
{\UCONN \small University of Connecticut, Storrs, Connecticut 06269} \\
{\ECOSSEE \small Edinburgh University, Edinburgh EH9 3JZ, United Kingdom} \\
{\EMMY \small Emmy-Noether Foundation, Germany} \\
{\FIU \small Florida International University, Miami, Florida 33199} \\
{\FSU \small Florida State University, Tallahassee, Florida 32306} \\
{\GWU \small The George Washington University, Washington, DC 20052} \\
{\ECOSSEG \small University of Glasgow, Glasgow G12 8QQ, United Kingdom} \\
{\ISU \small Idaho State University, Pocatello, Idaho 83209} \\
{\INFNFR \small INFN, Laboratori Nazionali di Frascati, Frascati, Italy} \\
{\ORSAY \small Institut de Physique Nucleaire ORSAY, Orsay, France} \\
{\BONN \small Institute f\"{u}r Strahlen und Kernphysik, Universit\"{a}t Bonn, Germany} \\
{\ITEP \small Institute of Theoretical and Experimental Physics, Moscow, 117259, Russia} \\
{\JMU \small James Madison University, Harrisonburg, Virginia 22807} \\
{\KYUNGPOOK \small Kyungpook National University, Daegu 702-701, South Korea} \\
{\MIT \small Massachusetts Institute of Technology, Cambridge, Massachusetts  02139-4307} \\
{\UMASS \small University of Massachusetts, Amherst, Massachusetts  01003} \\
{\UNH \small University of New Hampshire, Durham, New Hampshire 03824-3568} \\
{\NSU \small Norfolk State University, Norfolk, Virginia 23504} \\
{\OHIOU \small Ohio University, Athens, Ohio  45701} \\
{\ODU \small Old Dominion University, Norfolk, Virginia 23529} \\
{\PITT \small University of Pittsburgh, Pittsburgh, Pennsylvania 15260} \\
{\RPI \small Rensselaer Polytechnic Institute, Troy, New York 12180-3590} \\
{\RICE \small Rice University, Houston, Texas 77005-1892} \\
{\URICH \small University of Richmond, Richmond, Virginia 23173} \\
{\SCAROLINA \small University of South Carolina, Columbia, South Carolina 29208} \\
{\JLAB \small Thomas Jefferson National Accelerator Facility, Newport News, Virginia 23606} \\
{\UNIONC \small Union College, Schenectady, NY 12308} \\
{\VT \small Virginia Polytechnic Institute and State University, Blacksburg, Virginia   24061-0435} \\
{\VIRGINIA \small University of Virginia, Charlottesville, Virginia 22901} \\
{\WM \small College of William and Mary, Williamsburg, Virginia 23187-8795} \\
{\YEREVAN \small Yerevan Physics Institute, 375036 Yerevan, Armenia} \\
{\NOWOHIOU \small Ohio University, Athens, Ohio  45701} \\
{\NOWINDSTRA \small Systems Planning and Analysis, Alexandria, Virginia 22311} \\
{\NOWUNH \small University of New Hampshire, Durham, New Hampshire 03824-3568} \\
{\NOWMOSCOW \small Moscow State University, General Nuclear Physics Institute, 119899 Moscow, Russia} \\
{\NOWSCAROLINA \small University of South Carolina, Columbia, South Carolina 29208} \\
{\NOWUMASS \small University of Massachusetts, Amherst, Massachusetts  01003} \\
{\NOWMIT \small Massachusetts Institute of Technology, Cambridge, Massachusetts  02139-4307} \\
{\NOWURICH \small University of Richmond, Richmond, Virginia 23173} \\
{\NOWODU \small Old Dominion University, Norfolk, Virginia 23529} \\
{\NOWGEISSEN \small Physikalisches Institut der Universitaet Giessen, 35392 Giessen, Germany} \\

\end{center}
\end{titlepage}

The inclusive, inelastic $eD$ scattering cross section has
been measured with the CLAS~\cite{CLASNIM} detector
in Hall B of the Thomas Jefferson National Accelerator Facility (TJNAF).
Combining these data and previously measured world data we have extracted Nachtmann
moments~\cite{Nachtmann} of the deuteron structure function $F_2$ in the region
$0.4 < Q^2 < 100$~GeV$^2$/c$^2$.  These results are published in Ref.~\cite{mymoments}.
The purpose of the present CLAS-Note is to tabulate the CLAS deuteron $F_2$ data.
A description of the data analysis is reported in the Ref.~\cite{mymoments}.

The data were collected during the so called
``E1d'' and ``E6a'' electron beam running periods.
These runs were taken in March - April 2000 and January - March 2002, respectively.
The data were taken on the 5 cm long liquid deuteron (LD) target, located in
the center of CLAS. The trigger was set by a coincidence of
hits in the Cherenkov Counter (CC) and the Electromagnetic Calorimeter (EC).
In E6a run period Level 2 trigger (L2), requiring
a track candidate in the same sector of the EC hit, was set to reduce
the background rate~\cite{klim}.
More details on the run conditions and the total number of electron triggers collected
in these experiments are given in the Table~\ref{table:d_ad1}.

\begin{figure}[!h]
\begin{center}
\includegraphics[bb=2cm 4cm 22cm 24cm, scale=0.4]{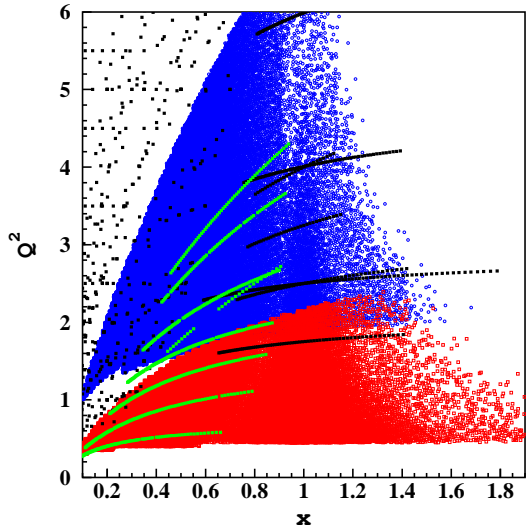}
\caption{Kinematic regions covered by measured data sets:
the red area represents the E1d run period data taken at the beam energy $E=2.474$ GeV;
the blue area indicates the E6a run period data taken at the beam energy $E=5.77$ GeV;
the green points represent measurements from Ref.~\cite{f2-hc};
the black points show measurements from Refs.~\cite{e133,e140,e140x,e49a,e49b,e61,e891,e892,ne11}.}
\label{fig:d_ad1}
\end{center}
\end{figure}

The kinematic region accessible with these data is shown in Fig.~\ref{fig:d_ad1}:
E1d run period data cover the lower $Q^2$ region from 0.4 up to 1.5 GeV$^2$,
while the higher $Q^2$ area is filled by E6a data, situated from 1.5 up to 6 GeV$^2$.
Combining the two data sets we obtained the continuous kinematic coverage from
0.4 up to 6 GeV$^2$ at large $x$ values.

\begin{table}[!h]
\begin{center}
\caption{Run conditions and the total number of triggers.}
\label{table:d_ad1}
\vspace{2mm}
\begin{tabular}{|c|c|c|c|c|} \hline
\multicolumn{4}{|c|}{Data set} & Events \\ \hline
Beam         & Torus       & Trigger & Beam         & x$10^6$ \\
energy [GeV] & current [A] &  [mV]   & current [nA] &  \\ \hline
2.474 & 1500 & CC(20)+EC(60-100)     & 2.5 & 150   \\ \hline
5.77  & 2250 & CC(100)+EC(72-172)+L2 & 8   & 1100  \\ \hline
\end{tabular}
\end{center}
\end{table}

The deuteron structure function $F^D_2(x,Q^2)$ was extracted from the inelastic
cross section using the fit of the function $R(x,Q^2)\equiv \sigma_L / \sigma_T$ as follows:
\begin{equation}
F^D_2(x,Q^2)=\frac{1}{\sigma_{Mott}}\frac{d^2\sigma}
{d\Omega dE^{\prime}}
\frac{\nu}{1+\frac{1-\epsilon}{\epsilon}\frac{1}{1+R}}
\label{eq:d_sf1}
\end{equation}
\noindent where the virtual photon polarization parameter is defined as
\begin{equation}
\epsilon \equiv \Biggl(
1+2\frac{\nu^2+Q^2}{Q^2} tan^2{\theta \over 2}
\Biggr )^{-1}
\label{eq:d_sf2}
\end{equation}

\begin{figure}[!h]
\centering
\subfigure[$Q^2=0.475$ GeV$^2$]{
\includegraphics[bb=2cm 4cm 18cm 24cm, scale=0.4]{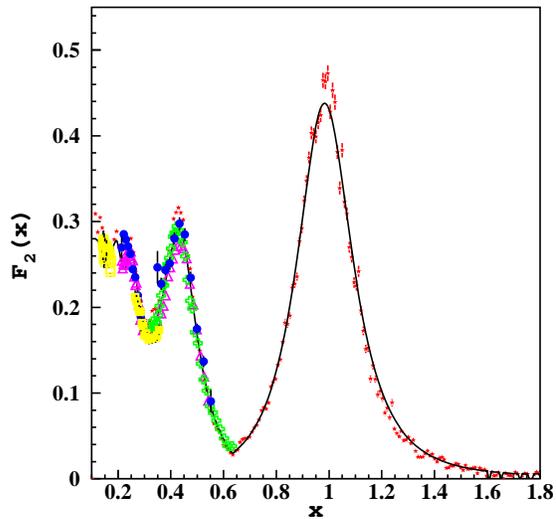}}~~~~~~~~
\subfigure[$Q^2=1.175$ GeV$^2$]{
\includegraphics[bb=2cm 4cm 18cm 24cm, scale=0.4]{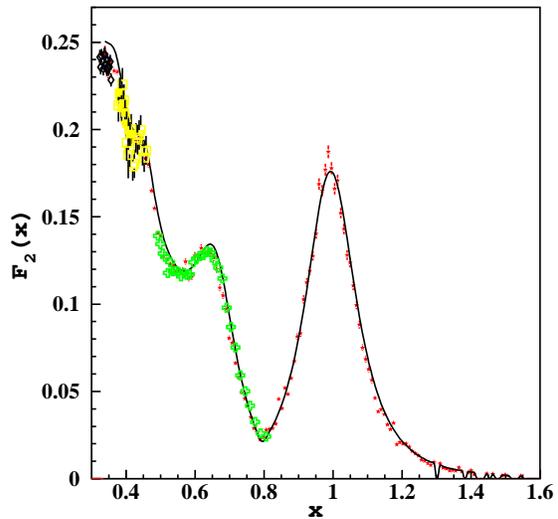}}
\subfigure[$Q^2=2.375$ GeV$^2$]{
\includegraphics[bb=2cm 4cm 18cm 24cm, scale=0.4]{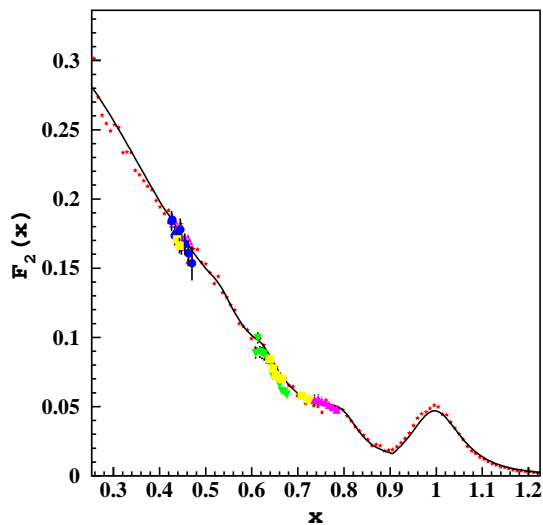}}~~~~~~~~
\subfigure[$Q^2=3.575$ GeV$^2$]{
\includegraphics[bb=2cm 4cm 18cm 24cm, scale=0.4]{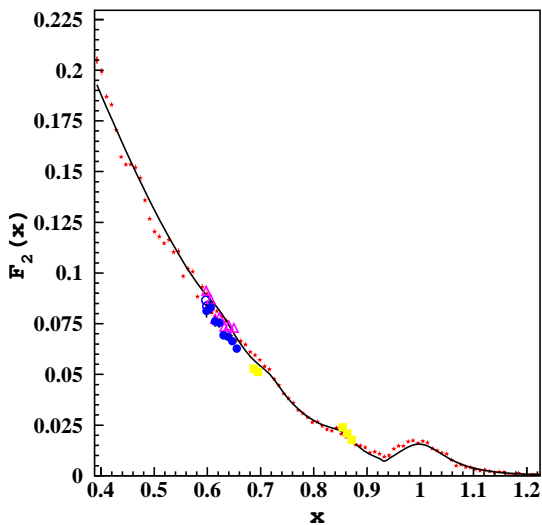}}
\caption{The deuteron structure function $F^D_2(x,Q^2)$ per nucleon for a few $Q^2$ values:
the red points show CLAS data obtained in the present analysis;
other color points represent various previous measurements from
Refs.~\cite{f2-hc,e133,e140,e140x,e49a,e49b,e61,e891,e892,ne11},
where each separate experiment is indicated by a different color;
the black solid line represents the phenomenological model used in the present
analysis and described in Ref.~\cite{mymoments}.}
\label{fig:r_ps1}
\end{figure}

\begin{figure}[!h]
\centering
\includegraphics[bb=2cm 4cm 18cm 24cm, scale=0.95]{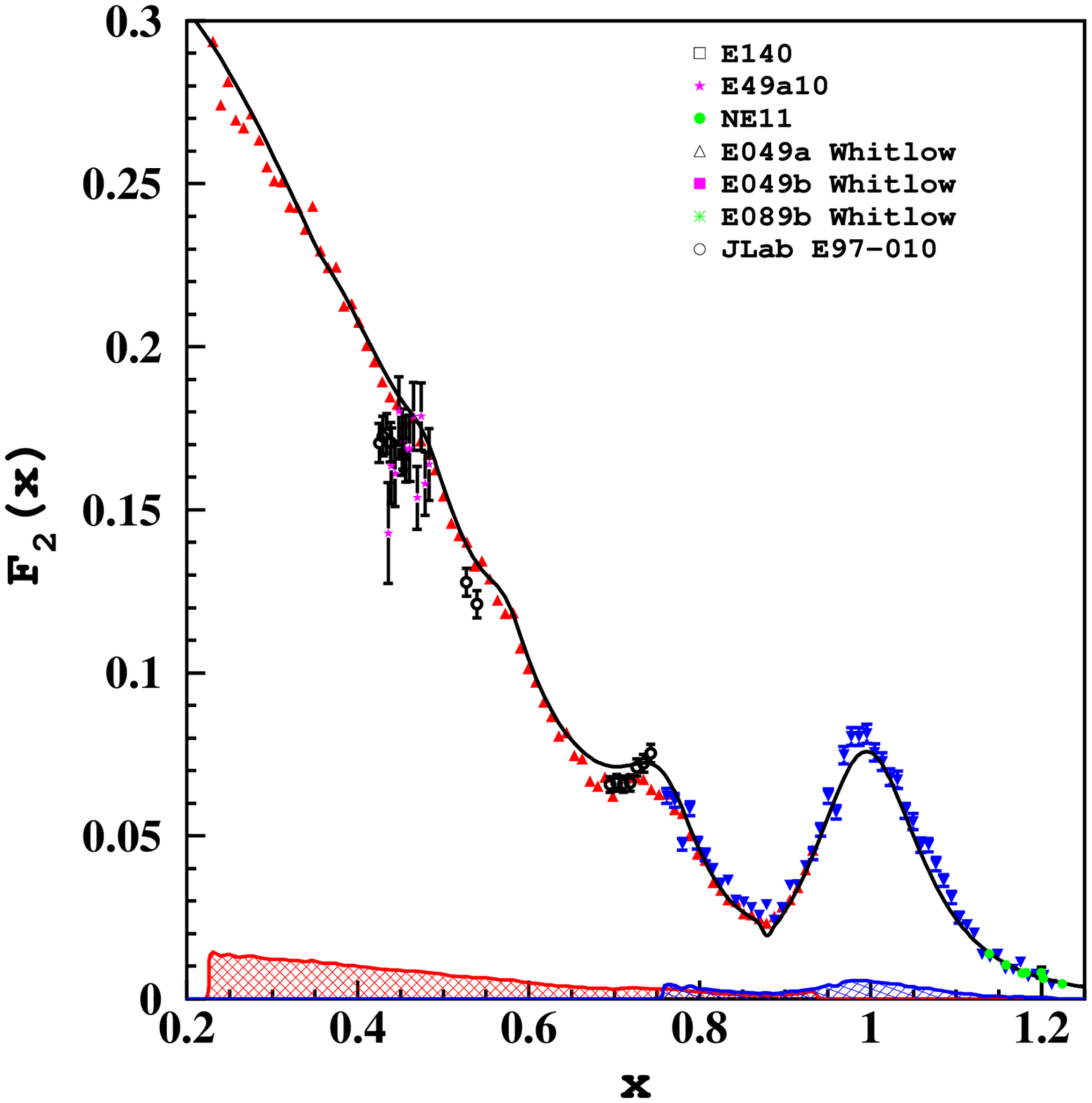}
\caption{The structure deuteron function $F^D_2(x,Q^2)$ per nucleon at $Q^2=1.925$ GeV$^2$:
the red and blue points show the CLAS data from E6a and E1d run periods, respectively;
other color points represent the previous measurements from
Refs.~\cite{f2-hc,e133,e140,e140x,e49a,e49b,e61,e891,e892,ne11}, where each separate
experiment is indicated by a different color;
the black solid line shows the phenomenological model used in the present analysis.}
\label{fig:r_ps5}
\end{figure}

\begin{figure}[!h]
\centering
\includegraphics[bb=2cm 4cm 18cm 24cm, scale=0.95]{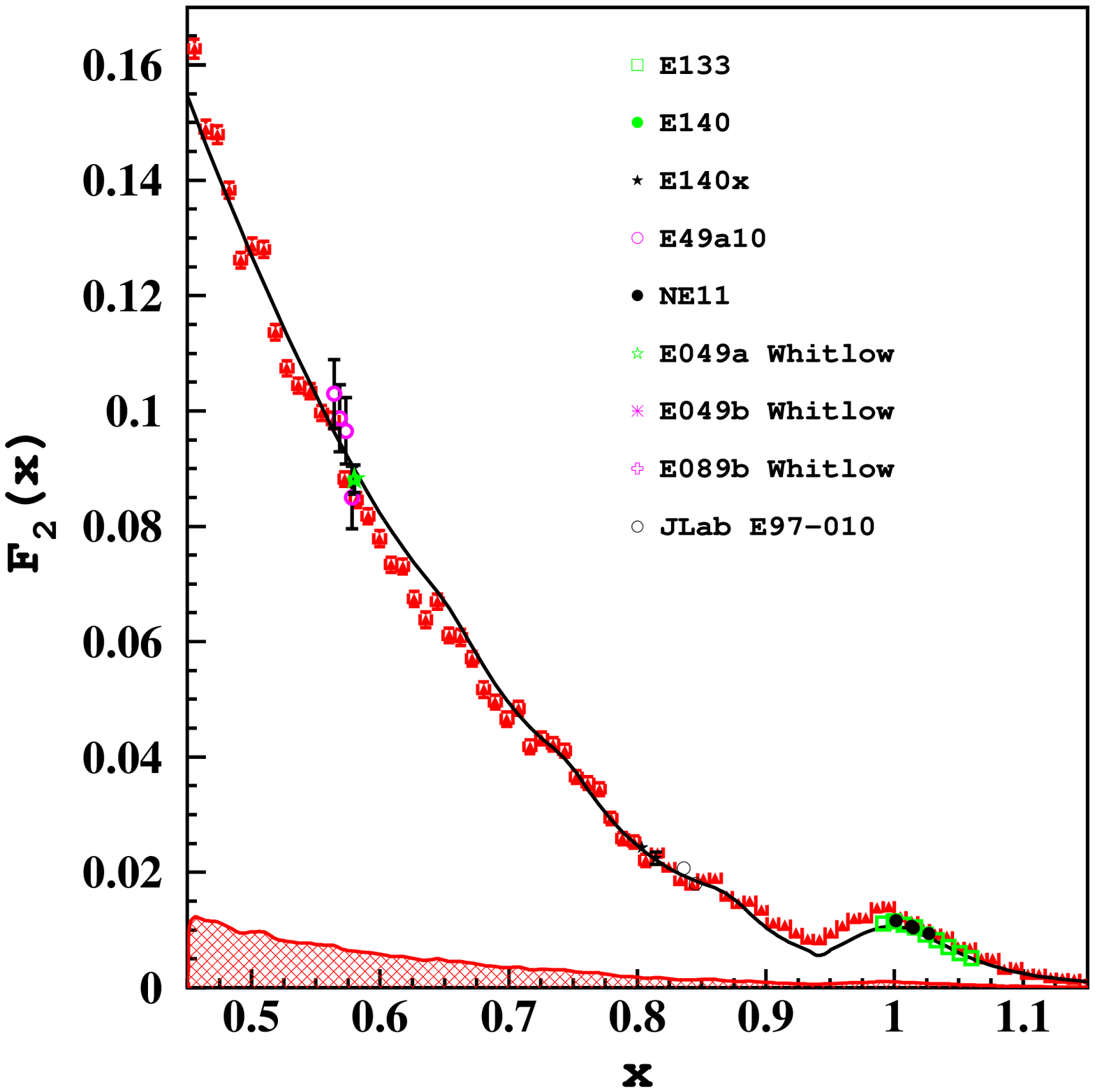}
\caption{The deuteron structure function $F^D_2(x,Q^2)$ per nucleon at $Q^2=4.075$ GeV$^2$:
the red and blue points show the CLAS data from E6a and E1d run periods, respectively;
other color points represent the previous measurements from
Refs.~\cite{f2-hc,e133,e140,e140x,e49a,e49b,e61,e891,e892,ne11}, where each separate
experiment is indicated by a different color;
the black solid line shows the phenomenological model used in the present analysis.}
\label{fig:r_ps7}
\end{figure}

The ratio $R(x,Q^2)$ for the proton is well established in the DIS region and can be
fairly well described by the SLAC parameterization from Ref.~\cite{r_fit_dis}. In the same time,
the experimental data in the resonance region were missing until recently.
The data from Hall C (TJNAF) published in Ref.~\cite{Keppel_R} cover the entire resonance
region and extend to very low $Q^2$ values. It was shown by HERMES
Collaboration in Ref.~\cite{HERMES_R} that in the DIS the ratio $R$ does not
depend on the nuclear mass number $A$. We take this assumption to be valid
also in the resonance region. But, since Fermi motion smearing effects
were expected to change both $F_2$ and $F_L$ structure functions
we performed a ``smooth'' parameterization of the measured ratio $R$.
In this way the resonance structures, clearly seen on the proton,
were averaged out in the medium curve. This ``smooth'' parameterization
was obtained by modifying the fit from Ref.~\cite{r_fit_dis} $R^{SLAC}$ as follows:
\begin{equation}
R^{smooth}=R^{SLAC}
\Biggl\{1-\frac{W_{TH}^2-C_W^2}{W^2}\Biggr\}^{B_W}
\Biggl(\frac{Q^2}{Q_0^2}\Biggr)^{C_Q} e^{-B_Q C_Q (Q^2/Q_0^2-1)}
\end{equation}
\noindent where $W$ is the invariant mass of the produced hadronic system,
$W_{TH}$ is its value at the pion threshold and the parameters are
found by a MINUIT minimization procedure:
$Q_0^2=0.8$ GeV$^2$,
$B_Q=2.14$,
$C_Q=0.729$,
$B_W=0.383$ and
$C_W=0.165$ GeV.

\begin{figure}[!h]
\centering
\includegraphics[bb=2cm 4cm 18cm 24cm, scale=0.4]{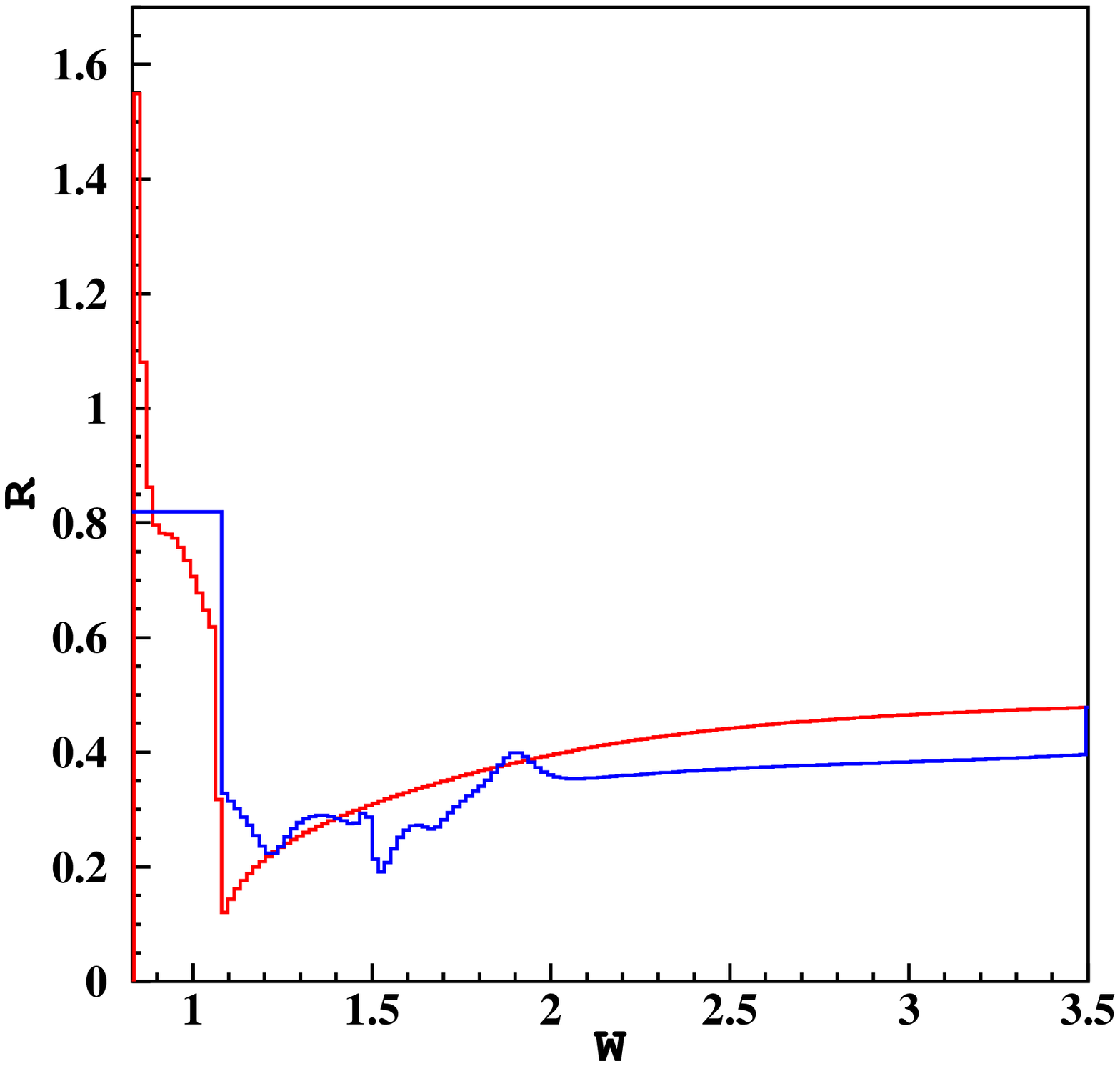}~~~~~~%
\includegraphics[bb=2cm 4cm 18cm 24cm, scale=0.4]{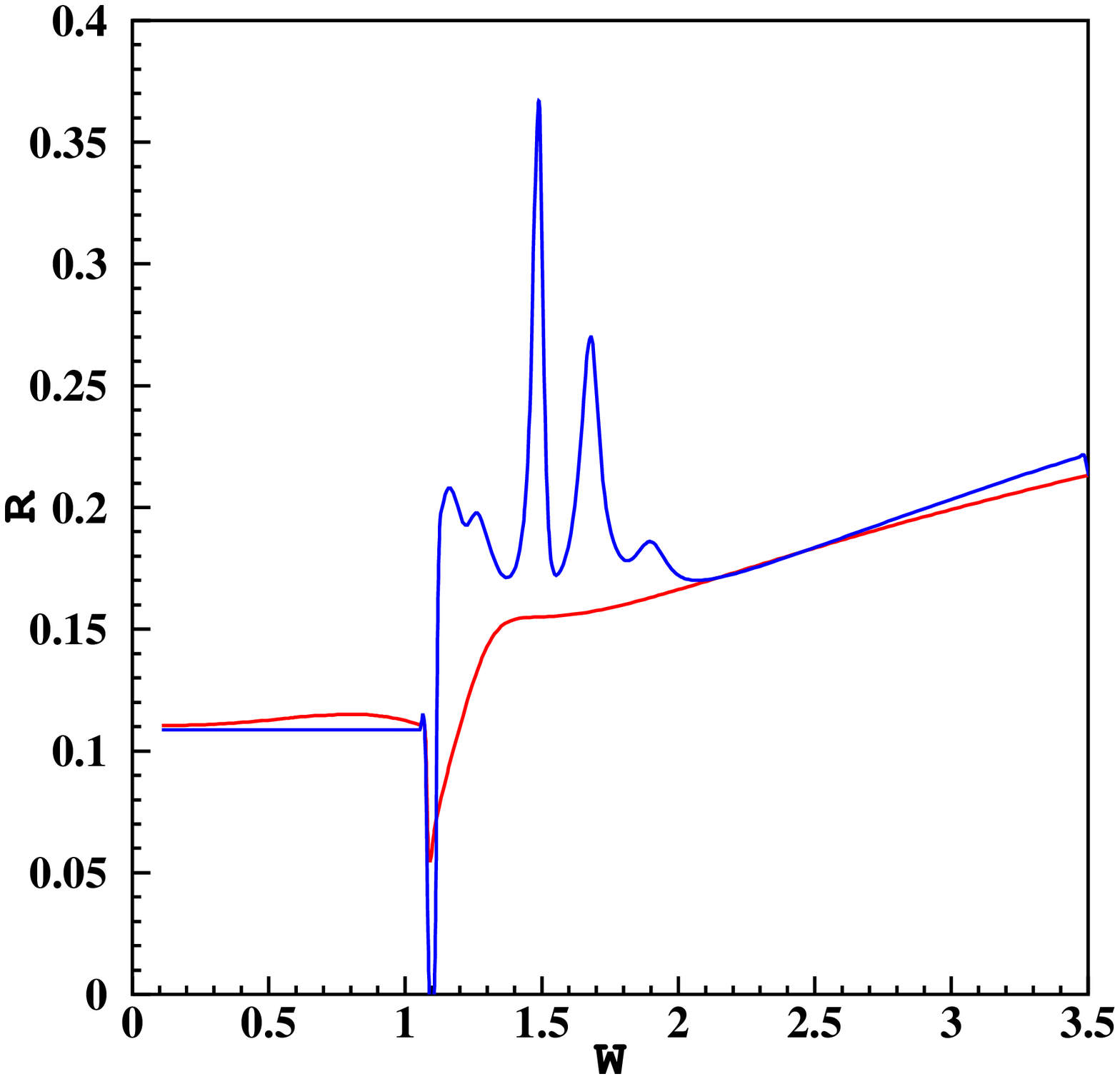}
\caption{The ratio $R=\sigma_L/\sigma_T$ as a function of the final hadronic system invariant mass $W$
at $Q^2=0.425$ GeV$^2$ (left) and at $Q^2=2.875$ GeV$^2$ (right):
the red curve represents a smooth fit to the world data in the inelastic region and
the model from Ref.~\cite{Simula_QE} in the quasi-elastic region;
the blue curve shows the fit from Ref.~\cite{Keppel_R} in the inelastic regime
and the direct sum of the nucleon form-factors under the quasi-elastic peak region.}
\label{fig:c_rlt1}
\end{figure}

The comparison of two models
is shown in Fig.~\ref{fig:c_rlt1}. The difference between these two models
gave us an estimate of the systematic errors of the ratio $R$, which turned
out to be small.

The ratio $R$ under the quasi-elastic peak is a separate issue. Because of
the quasi-elastic peak nature it is not more independent of $A$ and should be, therefore, treated
within a nuclear model framework. We used the model from Ref.~\cite{Simula_QE},
which treats separately the longitudinal and transverse nuclear response functions
$R_L$ and $R_T$. The conventional ratio $R$ can be calculated from these
quantities as follows:
\begin{equation}\label{eq:c_rlt1}
R=\frac{2Q^2}{Q^2+\nu^2}\frac{R_L}{R_T}
\end{equation}

\begin{figure}[!h]
\centering
\includegraphics[bb=2cm 4cm 18cm 24cm, scale=0.4]{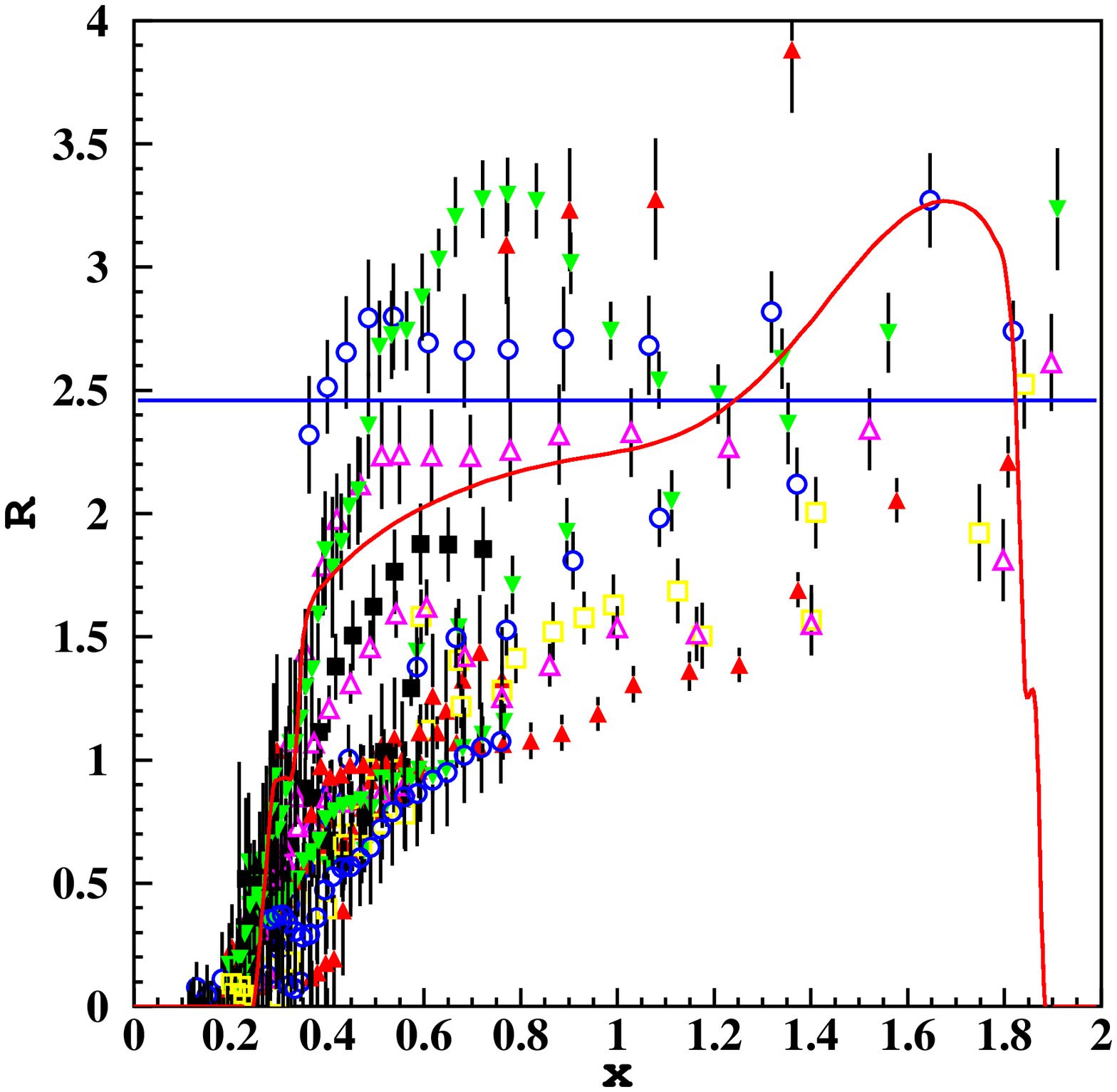}~~~~~~%
\includegraphics[bb=2cm 4cm 18cm 24cm, scale=0.4]{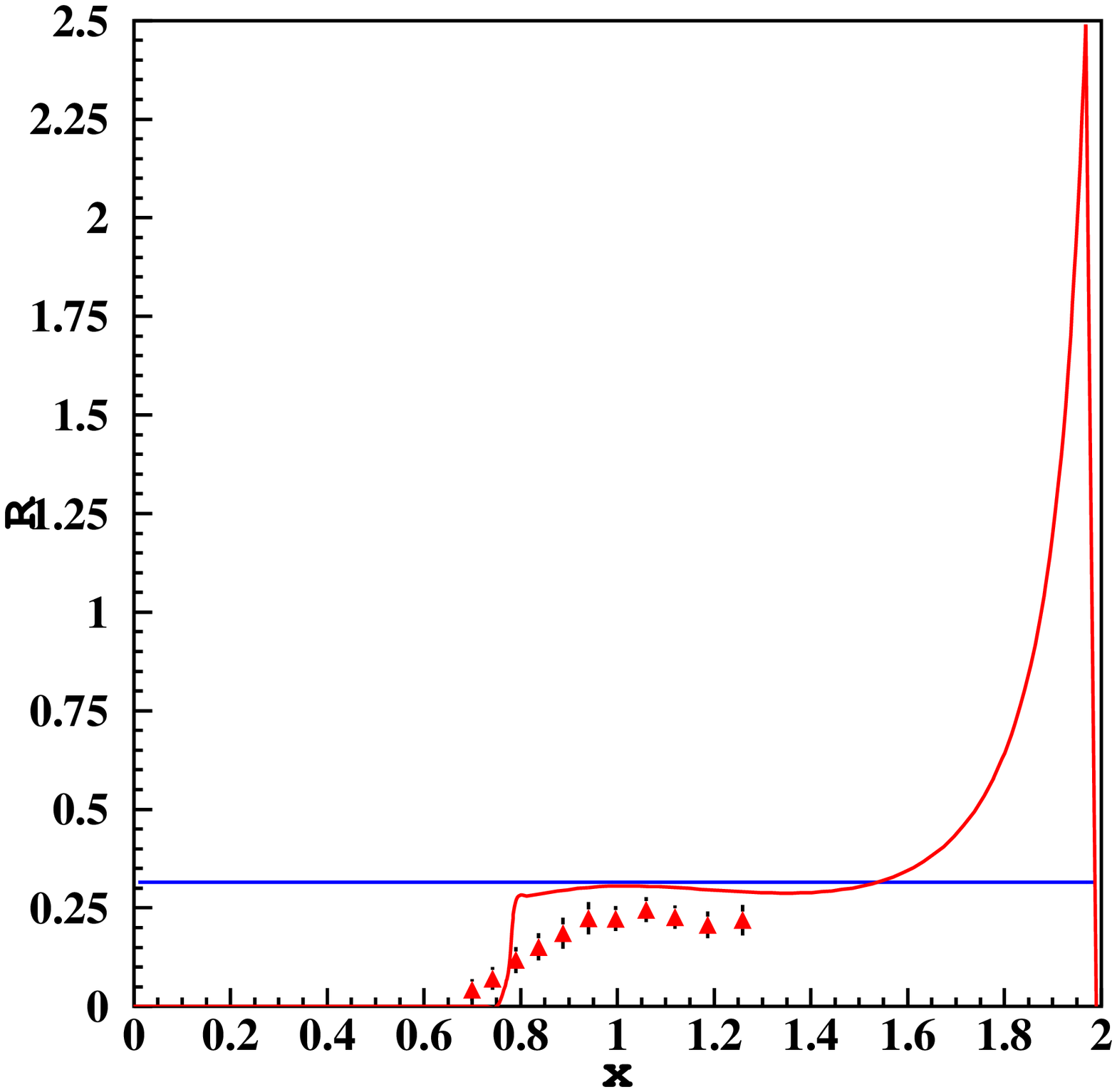}
\caption{The ratio $R=\sigma_L/\sigma_T$ in the quasi-elastic peak region as a function
of $x$ at $Q^2=0.1-0.15$ GeV$^2$ (left panel) and at $Q^2=0.85-1.1$ GeV$^2$ (right panel):
the red curve represents the model from the Ref.~\cite{Simula_QE};
the blue curve shows the sum of the nucleon form-factors;
the data points are taken from Refs.~\cite{r_qe_nucl}.}
\label{fig:c_rlt2}
\end{figure}

In the Fig.~\ref{fig:c_rlt2} the obtained deuteron quasi-elastic ratio $R$ is shown in comparison to the data
on the ratio $R$ for other nuclei (precise deuteron data on $R$ in the quasi-elastic peak region
are missing) and a sum of the proton and neutron form factors, which simply implies:
\begin{equation}\label{eq:c_rlt2}
R=\frac{G_E^2}{\tau G_M^2}
\end{equation}
\noindent where $\tau=Q^2/4M^2$ and $G_E$, $G_M$ are the sums of the known Sachs form-factors of
the proton and neutron. Since the number of protons and neutrons is different
in different nuclei we did not expect to have the same ratio $R$ for all of them.
However, the $x$-shape of the ratio is likely very similar from nucleus to nucleus.
In fact, in Fig.~\ref{fig:c_rlt2} it can be clearly seen. The calculations
agrees reasonably well with the $x$-shape of the data, while the absolute values
can be different.
The systematic error can be estimated as a difference between the model calculation
and the result of the naive form-factor sum through Eq.~\ref{eq:c_rlt2}.

\FloatBarrier

\setlength\textheight{750pt}
\setlength\voffset{-0.5in}
\setlength\topmargin{1pt}
\setlength\headheight{1pt}
\setlength\headsep{1pt}
\setlength\footskip{0pt}
\setlength\marginparwidth{1pt}
\renewcommand{\topfraction}{0.99}
\renewcommand{\bottomfraction}{0.99}


\begin{center}
\twocolumn
\topcaption{$F_2(x,Q^2)$ with statistical and systematic errors
extracted from CLAS data.}\label{tab:f2_clas}
\scriptsize
\tablehead{\hline}
\tabletail{\hline}
\begin{xtabular}{|c|c|c|c|c|} \hline
$Q^2$     & $x$ & $F_2$ & $\delta_{stat}$ & $\delta_{sys}$ \\ \hline
[GeV$^2$] &     &       &                 &                \\ \hline
0.3750 & 0.1135 &  2.6899E-01 &  7.7614E-04 &  1.1663E-02 \\ \cline{2-5}
 & 0.1225 &  2.6095E-01 &  1.0030E-03 &  1.1188E-02 \\ \cline{2-5}
 & 0.1315 &  2.6306E-01 &  1.3176E-03 &  1.1238E-02 \\ \cline{2-5}
 & 0.1405 &  2.5496E-01 &  1.5563E-03 &  1.0820E-02 \\ \cline{2-5}
 & 0.1495 &  2.6625E-01 &  1.8788E-03 &  1.1302E-02 \\ \cline{2-5}
 & 0.1585 &  2.7369E-01 &  2.1750E-03 &  1.1614E-02 \\
\hline
\end{xtabular}
\end{center}

\end{document}